\documentstyle[12pt]{article}
\begin{document}
\title{ Constrained Euler-Poincar\'{e} Supergravity \\ in Five Dimensions}
\author { E . Atasoy  \\
{\small Department of Physics Engineering }\\
{\small Hacettepe University} \\
{\small 06532 Ankara, Turkey} \and
T. Dereli \thanks {E.mail: tekin@dereli.physics.metu.edu.tr} \\
{\small Department of Physics} \\
{\small Middle East Technical University} \\
{\small 06531 Ankara, Turkey  } \and
M . \"{O}nder \thanks {E.mail: Fonder@eti.cc.hun.edu.tr}\\
{\small Department of Physics Engineering }\\
{\small Hacettepe University} \\
{\small 06532 Ankara, Turkey} }
\date{July 29, 1996}
\maketitle
\begin{abstract}
The N=2 supergravity action in D=5
is generalized by the inclusion of  dimensionally continued
Euler-Poincar\'{e} form. The spacetime torsion  
implied by the Einsteinean supergravity  is
imposed by a Lagrange constraint and the resulting
variational equations are solved for the Lagrange multipliers. 
The corresponding terms in the Einstein and
Rarita-Schwinger field equations are determined. 
These indicate new types of interactions that 
could be included in the action to achieve local supersymmetry.
\end{abstract}
\newpage
\noindent {\bf 1. Introduction}
\vskip 2mm
The long ranged interactions of nature can be formally unified with
gravity in spacetimes of dimensions greater than
four. The Einstein-Hilbert action is usually taken as the basis for
this kind of unification. In this case the zero-torsion constrained variations
of the gravitational action yields the Einstein tensor that
has the property of being covariantly constant and
involving at most the second order partial derivatives of the metric tensor
components. These properties are unique to the Einstein tensor for 
spacetime dimension D=4.
However, for $D>4$ the tensors obtained from the 
zero torsion constrained variations of
dimensionally continued Euler forms all share these properties.
Therefore we contemplate unified theories in  spacetime dimensions
$D > 4$  with a gravitational action that
is a linear combination of all dimensionally continued Euler-Poincar\'{e}
forms including the Einstein-Hilbert term [1],[2].
Fermions can be incorporated in such unified models by 
requiring local supersymmetry. Thus  
it seems natural to ask for an Euler-Poincar\'{e} supergravity
in $D >4$, however, locally supersymmetric extension
of the dimensionally continued Euler forms is not easy to construct.
The simplest model  that we can use for this kind of 
generalisation is provided by N=2 supergravity in D=5 dimensions.
 
The kinematics of $D=5$ spinors  given by Cremmer [3] are used by
Chamseddine and Nicolai [4] to construct the Einsteinean supergravity
action. The same theory is constructed
independently  by
D'Auria and Fr\'{e} [5] using the group manifold approach.
The construction of the Euler-Poincar\'{e} supergravity using 
the group manifold approach
is discussed by Ferrara, Fr\'{e} and Porrati [6].
The Noether construction of
Euler-Poincar\'{e} supergravity is  taken by Ro\u{c}ek, van
Nieuwenhuizen and Zhang [7]. Both these approaches yield  only partial results so that a complete action
with local supersymmetry is not yet available. 
Even at this level the classical solutions of the Euler-Poincar\'{e}
supergravity theory show some interesting features [8],[9],[10].

Here we wish to  offer some additional
understanding derived from the techniques of constrained variations [11].
In the case of Einsteinean supergravity the independent connection
variations of the action yield a set of field equations that can be solved
algebraically, thus determining the spacetime torsion in terms of
expressions that are quadratic in gravitino fields.
It is well known that the same theory is obtained under the zero-torsion
constraint (i.e. spacetime is pseudo-Riemannian or the connection is
Levi-Civita) provided appropriate quartic gravitino self-interactions are
included in the action to guarantee local supersymmetry.
The situation changes drastically when Euler-Poincar\'{e} gravity is
considered. In this case, when the metric and the connection are varied
independently, the connection variation equations contain
both the curvature and the torsion tensors explicitly. 
Then it is not possible to
express the spacetime torsion solely in terms of the gravitino fields.
A way of approach is to accept this situation as it is, treating
torsion as a true dynamical degree of freedom, and to search for a locally
supersymmetric action.
The other avenue of approach is to constrain the torsion  to
some desired expression in terms of other field variables, and 
implement this constraint by the method of Lagrange multipliers. 
In the following we constrain 
the spacetime torsion to whatever it is in the
Einsteinean supergravity.
We solve connection variation equations for the Lagrange multiplier forms
and substitute these into the Einstein 
and  Rarita-Schwinger equations. Thus we are
able to delineate new non-linear interactions 
implied by our
torsion constraint. Whether these will be relevant to the construction
of a locally  supersymmetric 
Euler-Poincar\'{e} supergravity action remains to be seen.
\vskip 2mm
\noindent {\bf Notation and Conventions}
\vskip 2mm
\noindent The minimal supergravity multiplet in D=5 dimensions contains 

i) The metric tensor of spacetime
\begin{equation}
g = \eta_{AB} e^{A} \otimes e^{B}
\end{equation}
where we take the spacetime metric with signature $\eta_{AB}$ = diag$(-++++)$
and coframe 1-forms ( e$^{A}$ ) are dual to the orthonormal frame vectors
($X_{A}$)    so that g($X_{A},X_{B}$) = $\eta_{AB}$.
A, B,...= 0, 1, 2, 3, 5 are frame indices.

ii) so(2)-valued gauge potential 1-form $iA$ 
is introduced to complete the bosonic
degrees of freedom. The corresonding gauge field 2-form is 
\begin{equation}
F=dA.
\end{equation}
iii) The fermionic degrees of freedom are carried by the
symplectic Majorana spinor valued 1-forms
\begin{equation}
\psi ^{I} = \psi ^{I}_{A} e^{A},  \hspace*{1cm} I = 1, 2
\end{equation}

We will  exploit the isomorphism between the Clifford algebra $Cl(1,4)$ and
total matrix algebra ${\cal M}_4$ [12] so that the Clifford algebra generators
$\{\Gamma _{A}\}$ are realised by a set of $4\times4$ matrices
that satisfy
\begin{equation}
\Gamma _{A} \Gamma _{B} + \Gamma _{B} \Gamma _{A} = 2\eta _{AB}I.
\end{equation}
With our conventions $\Gamma _{0}$  is anti-Hermitean, while the remaining
generators  $\Gamma _{1}, \Gamma _{2}, 
\Gamma _{3}, \Gamma _{5}$
are Hermitean. In D=5 we have $\Gamma _{0} \Gamma _{1} \Gamma _{2}
\Gamma _{3} = i \Gamma _{5}$
which implies the following identities: 
$$\Gamma _{AB} = \frac{1}{3!} \epsilon _{ABCDF} \Gamma ^{CDF},$$
$$\Gamma _{ABC} = \frac{1}{2!} \epsilon _{ABCDF} \Gamma ^{DF},$$
$$\Gamma _{ABCD} = \epsilon _{ABCDF} \Gamma ^{F},$$
\begin{equation}
\Gamma _{ABCDF} = \epsilon _{ABCDF} I.
\end{equation}
For the construction of spinors in $D=5$ 
we keep close  to the definitions of Cremmer [3] 
and let symplectic Majorana spinors be given by
$$\psi^{I} = C_{5} ({\bar {\psi}}^{I})^{T} $$
where the charge conjugation matrix  satisfies
$$ C_{5}\Gamma _{A} C_{5}^{-1} = \Gamma _{A}^{T}.$$
We may take  $ C_{5} = \Gamma _{0}\Gamma _{5}$ so that
$\psi ^{I} = \Gamma _{5} ({\psi ^{I}})^{*}$, \hspace*{4mm}I = 1, 2.
With the above definitions all the Majorana flip identities 
can be encoded into the single expression
\begin{equation}
\bar{\psi ^{I}} \Gamma _{A_{1}}\Gamma _{A_{2}} \cdots \Gamma _{A_{k}} \phi ^{J}
= \bar{\phi ^{J}}\Gamma _{A_{k}} \cdots \Gamma _{A_{2}} \Gamma _{A_{1}} \psi^{I}
, \hspace*{4mm} 0\leq k\leq 5.
\end{equation}
We raise or lower symplectic indices I, J = 1, 2 by the $2 \times 2$ 
matrix
$$(\epsilon)_{IJ} =  \left( \begin{array}{cc}
         0 & 1 \\
        -1 & 0 \\
\end{array} \right )$$
so that e. g.  $\psi _{I} = \epsilon _{IJ} \psi ^{J} $.   
Finally we note the identity
$$\bar{\psi ^{I}} M \phi _{I} = - \bar{\psi _{I}} M \phi ^{I}$$
\vskip 4mm
\noindent {\bf 2. Einsteinean Supergravity }
\vskip 3mm
The Einsteinean supergravity is described by a variational principle from
the action $ I_{o} = \int _{M} L_{o} $
where the Lagrangian 5-form
\begin{equation}
L_{o} = \frac{1}{2} R_{AB} \wedge ^{\star} (e^{A} \wedge e^{B}) +
\frac{i}{2} \bar{\psi ^{I}}^{\star}(\Gamma \wedge \Gamma \wedge
\Gamma)\wedge D\psi _{I} + \frac{1}{2} F\wedge ^{\star} F + \frac{k}{3}
F\wedge F\wedge A + \cdots
\end{equation}
In the above expression $\star : \Lambda ^{p} (M) \rightarrow \Lambda
^{5-p} (M)$ is the Hodge map on the algebra of exterior forms, determined by
the orientation $\star 1 = e^{o} \wedge e^{1} \wedge e^{2} \wedge e^{3}
\wedge e^{5}$. The exterior covariant derivative of a spinor
\begin{equation}
D\psi^{I} = d\psi^{I} + \Omega \wedge \psi^{I} 
\end{equation}
where the metric compatible connection 1-form
$$ \Omega = \frac{1}{2} \Omega ^{AB} \Sigma _{AB}$$
with the $so(4,1)$ algebra generators  $$ \Sigma _{AB} =
\frac{1}{4}[\Gamma _{A}, \Gamma _{B}] = \frac{1}{2} \Gamma _ {AB}.$$
Then we have
\begin{equation}
D^{2} \psi^{I} = \frac{1}{2} R^{AB} \Sigma _{AB} \wedge \psi^{I}
\end{equation}
where the curvature 2-forms
$$ R^{A}_{\hspace{1mm}B} = d\Omega ^{A}_{\hspace{1mm}B} + 
\Omega ^{A}_{\space{1mm}C} \wedge \Omega ^{C}_{\hspace{1mm}B}.$$
We have $\Gamma = \Gamma_{A} e^{A}$ so that using $ \Gamma$ - matrix
identities and the properties of the Hodge map we find
$$^{\star}(\Gamma \wedge \Gamma \wedge \Gamma ) = \Gamma \wedge \Gamma.$$
This identity allows us to simplify the Rarita-Schwingewr action written above.
But it should be remembered that this simplification can be done only in $D=5$.
In the Lagrangian above, we left terms that are needed for establishing
local supersymmetry; namely terms of the generic types 
$(\bar{\psi}\psi)^{2}$ and $(\bar{\psi}\psi F)$.
These are not essential for the arguments that follow.
Infinitesimal local supersymmetry transformations
$$\delta_{s} e^{A} = i\bar{\epsilon}^{I} \Gamma^{A}\psi_{I}$$
$$\delta_{s} \psi^{I} = 2D\epsilon^{I} + \cdots$$
\begin{equation}
\delta_{s} A = i\bar{\epsilon}^{I} \psi_{I}
\end{equation}
change $L_{o}$ by a closed form, so that the action remains invariant. 
Again we have left out terms
from the gravitino variations, of the generic type $(\bar{\epsilon}\psi) F$.
The connection is to be varied independently of the metric. Let us here
concentrate on the corresponding variational field equations :
\begin{equation}
^{\star}e_{ABC} \wedge T^{C} =
\frac{i}{4}\bar{\psi}^{I}\wedge\Gamma\wedge\Gamma
\wedge\Gamma_{AB}\psi_{I}.
\end{equation}
Then using the $\Gamma$-matrix identity
$$\Gamma_{CD}\Gamma_{AB} = (\eta_{CA}\eta_{DB} - \eta_{CB}\eta_{DA})I +
\epsilon_{CDABF} \Gamma^{F}$$
we solve for the torsion 2-forms as
\begin{equation}
T^{A} = \frac{i}{4}\bar{\psi}^{I}\wedge\Gamma^{A}\psi_{I} + \frac{i}{2}
\iota^{A} {^{\star}(\bar{\psi}^{I}\wedge\psi_{I})}.
\end{equation}
\vskip 4mm
\noindent {\bf 3. Euler-Poincar\'{e} Supergravity} 
\vskip 3mm
Now we are ready to consider the contribution of the Euler-Poincar\'{e} 
Lagrangian 5-form
\begin{equation}
L_{1} = \frac{k}{2}R_{AB}\wedge R_{CD}\wedge ^{\star}e^{ABCD}
\end{equation}
where $k$ is a coupling constant.
If $\int_{M}(L_{o} + L_{1})$ is varied
with respect to the connection, the Euler-Poincar\'{e} Lagrangian contributes
a term that involves both curvature and torsion explicity.
In this case it would not be possible to give the torsion
algebraically by an expression that is quadratic in the gravitino fields .
This will pose technical problems when one tries to establish a locally
supersymmetric extension. We wish to constrain the spacetime torsion to
what we already have in the Einsteinean supergravity. Then we will consider 
the constrained variations of the action and 
will be able to delineate some new 
interactions thus implied. To this end
we introduce the constraint Lagrangian 5-form
\begin{equation}
L_{constraint} = \lambda_{A}\wedge (de^{A} + \Omega^{A}_{\hspace{1mm}B}\wedge e^{B}
- \frac{i}{4}\bar{\psi}^{I}\wedge \Gamma^{A}\psi_{I} -
\frac{i}{2}{\iota^{A}}^{\star}
(\bar{\psi}^{I}\wedge \psi_{I}))
\end{equation}
where $\lambda_{A}$ are Lagrange multiplier 3-forms. Then the connection
variation of the total action
\begin{equation}
\int_{M} ( L_{o} + L_{1} + L_{constraint} )
\end{equation}
yields 
\begin{equation}
\lambda_{A}\wedge e_{B} - \lambda_{B}\wedge e_{A} = 4k\epsilon_{ABCDF} R^{CD}
\wedge T^{F}.
\end{equation}
We define tensor valued 3-forms
\begin{equation}
M_{AB} = \epsilon_{ABCDF} R^{CD}\wedge T^{F}
\end{equation}
and contract (16) by the interior product operators $\iota^A$ to get
\begin{equation}
\lambda_{B} + (\iota^{A} \lambda_{A}) \wedge e_{B} = 4k \iota^{A} M_{AB}.
\end{equation}
Contracting the above expression once again by $\iota^{B}$
we can solve (16) for the Lagrange multipliers and express them as
\begin{equation}
\lambda_{A} = 4k \iota^{B} M_{BA} - k e_{A}\wedge (
\iota^{B}\iota^{C} M_{CB}).
\end{equation}
Now, we go back to the coframe variations of the total action. 
From the first two terms we obtain the contributions
\begin{equation}
\frac{1}{2} R^{BC}\wedge^{\star} e_{ABC} + \frac{k}{2} R^{BC}\wedge R^{DF}
\wedge^{\star} e_{ABCDF} + \tau_{A} [\psi] + \tau_{A} [F] + \dots
\end{equation}
where $$\tau_{A}[\psi] = -i \bar{\psi^I} \wedge  \Gamma_{A} \Gamma \wedge 
D{\psi_{I}}$$ 
and 
$$\tau_{A}[F] = \frac{1}{2}(\iota_{A} F \wedge \star F - F \wedge \iota_{A} \star F)$$ are the energy-momentum
4-forms of the gravitino and gauge fields respectively. On the other hand
the variation of the constraint Lagrangian gives
\begin{equation}
D\lambda_{A} + \frac{1}{2} e_{A}^{\hspace{2mm}BCD}\wedge\lambda_{D} (\bar{\psi}^{I}_{B}\psi_{CI})
+ (\iota_{B}\lambda_{D})\wedge^{\star} e^{DBC} (\bar{\psi}^{iI}_{A}\psi_{CI}).
\end{equation}
Thus we see that through the Lagrange multipliers
Einstein equations get modified by terms that depend
explicitly on the
curvature of spacetime [13].  
Furthermore there are direct gravitino-curvature coupling terms of the type
$ (\bar{\psi}D\psi)R$  
and some quartic gravitino terms like 
 $R(\bar{\psi}\psi)^{2}$ that 
are implied by the quadratic gravitino terms present in the 
Lagrange multipliers (18).
To complete the discussion we also
show the modification to the Rarita-Schwinger equation:
$$^{\star}(\Gamma\wedge\Gamma\wedge\Gamma)\wedge D\psi^{I} + T^{A}\wedge
e^{B}\Gamma_{AB}\wedge\psi^{I} - \frac{i}{2}\lambda_{A}\wedge\Gamma^{A}
\psi^{I} + i ^{\star}(\iota^{A}\lambda_{A})\wedge\psi^{I} = 0 $$
Again terms that  involve the curvature and also gravitino
self-couplings are generated by the corresponding terms  
implicit in $ \lambda _{A}$`s.
\vskip 4mm
\noindent {\bf 4. Conclusion}
\vskip 4mm
In this work some properties of the
Euler-Poincar\'{e} supergravity in $D=5$ dimensions have been studied.
The spacetime torsion is constrained to its form in
Einsteinean supergravity  
by the method of Lagrange multipliers. Then the connection variation 
equations are solved for the Lagrange multiplier forms. 
These are inserted into the
Einstein and Rarita-Schwinger field equations that are obtained through 
the co-frame and gravitino variations, respectively. Thus new types of 
interactions are exhibited. In particular, Einstein equations contain
terms of the types $(\bar{\psi} \psi) R$ and $R (\bar{\psi} \psi)^2$.
The new terms in the Rarita-Schwinger equation involve
both curvature and gravitino self couplings.
It is suggested that these new types of interactions may be introduced in the
action to achieve local supersymmetry of the 
Euler-Poincar\'{e} supergravity.
\newpage

\noindent {\bf References}
\vskip 4mm
\begin{description}
\item[[1]] M. Ar{\i}k, T. Dereli, Phys.Lett.{\bf B189}(1986)96 
\item[[2]] M. Ar{\i}k, T. Dereli, Phys. Rev. Lett. {\bf 63}(1989)5
\item[[3]] E. Cremmer {\bf Superspace and Supergravity}
Edited by S. W. Hawking, M. Ro\u{c}ek (Cambridge U.P., 1981) p.267
\item[[4]] A. Chamseddine, H. Nicolai, Phys.Lett.{\bf B96}(1980)89
\item[[5]] A. D'Auria, P. Fr\'{e}, Nucl. Phys.{\bf B201}(1982)101
\item[[6]] S. Ferrara, P. Fr\'{e}, M. Porrati, Ann. Phys. {\bf 175}(1987)112
\item[[7]] M.Ro\u{c}ek, P.van Nieuwenhuizen, S. C. Zhang,   
Phys.Rev.{\bf D33}(1986)370
\item[[8]] J. C. Fabris, R. Kerner, Helv. Phys. Acta{\bf 62}(1989)427
\item[[9]] R. Balbino, J. C. Fabris, R. Kerner, Phys.Rev.{\bf D42}(1990)1023
\item[[10]] R. Balbino, J. C. Fabris, R. Kerner, Class. Q. Grav.{\bf 7}(1990)L17
\item[[11]]  T. Dereli, R. W. Tucker, Nucl. Phys. {\bf B209}(1982)217
\item[[12]] I. M. Benn, R. W. Tucker{ \bf Introduction to Geometry and Spinors}
\\(Adam Hilger, 1989)
\item[[13]] E. Atasoy, Unpublished Ph. D. Thesis (Hacettepe University, Ankara, 1996)
\end{description}

\end{document}